\documentclass[conference]{IEEEtran}
\IEEEoverridecommandlockouts
\usepackage{cite}
\usepackage{amsmath,amssymb,amsfonts,amsthm}
\usepackage[ansinew]{inputenc}
\usepackage{algorithmic}
\usepackage{algorithm}
\usepackage{graphicx}
\usepackage{textcomp}
\usepackage{xcolor}
\usepackage{comment}
\usepackage{mathtools}
\usepackage[export]{adjustbox}
\def\BibTeX{{\rm B\kern-.05em{\sc i\kern-.025em b}\kern-.08em
    T\kern-.1667em\lower.7ex\hbox{E}\kern-.125emX}}
\begin{document}
\title{Event-Based Analysis of Solar Power Distribution Feeder Using Micro-PMU Measurements}

\author{\IEEEauthorblockN{
Parviz Khaledian, 
Armin Aligholian,
and Hamed Mohsenian-Rad 
\thanks{
The authors are with the Department of Electrical and Computer Engineering, University of California, Riverside, CA, USA. 
This work was supported in part by UCOP grant LFR-18-548175 for conducting the analysis and by DoE grant DE-EE8001 for obtaining the raw measurements. The corresponding author is H. Mohsenian-Rad; e-mail: hamed@ece.ucr.edu.}
}

\IEEEauthorblockA{Department of Electrical and Computer Engineering, University of California, Riverside, CA, USA}\vspace{-0.5cm}}

\maketitle

\begin{abstract}

Solar distribution feeders are commonly used in solar farms that are integrated into distribution substations. 
In this paper, we focus on a real-world solar distribution feeder and conduct an \emph{event-based analysis} by using micro-PMU measurements. The solar distribution feeder of interest is a behind-the-meter solar farm with a generation capacity of over 4 MW that has about 200 low-voltage distributed photovoltaic (PV) inverters. The event-based analysis in this study seeks to address the following practical matters.
First, we conduct event detection by using an unsupervised machine learning approach. For each event, we determine the event's source region by an impedance-based analysis, coupled with a descriptive analytic method. We segregate the events that are caused by the solar farm, i.e., \emph{locally-induced} events, versus the events that are initiated in the grid, i.e., \emph{grid-induced} events, which caused a \emph{response} by the solar farm.
Second, for the locally-induced events, we examine the impact of solar production level and other significant parameters to make statistical conclusions.  
Third, for the grid-induced events, we characterize the response of the solar farm; and  make comparisons with the response of an \emph{auxiliary neighboring feeder} to the same events. Fourth, we scrutinize multiple specific events; such as by revealing the dynamics to the control system of the solar distribution feeder. The results and discoveries in this study are informative to utilities and solar power industry.

\end{abstract}

\vspace{0.3cm}


\section{Introduction}\label{introduction}
As the penetration of solar power generation continues to grow, system operators confront new challenges.
Some of these challenges are introduced either by the power system \emph{events} which have impact on solar farms; or by locally generated \emph{events} that cause power quality issues due to the sharp drop and spike in solar power production \cite{jain2018power}.

To recognize these events and analyze their signatures and impacts, micro-PMU measurements can be of great value; given their high reporting rate of 120 phasor measurements per second; and their synchronization capability. Availability of such data has enabled high-resolution event analysis by using data mining and machine learning techniques \cite{ aligholian2019event,situational}. 

\vspace{0.05cm}

\subsection{Approach and Scope of Analysis}

The study in this paper is about a real-world solar distribution feeder that is integrated into a distribution substation. This solar distribution feeder is a large behind-the-meter solar farm, which is monitored by a micro-PMU at the distribution substation. 
We conduct an \emph{event-based analysis} of the micro-PMU measurements and report and discuss our discoveries. 

The events are first detected by an unsupervised machine learning method.
Next, we conduct the following analysis: 

\begin{enumerate}
    \item For each captured event, we seek to first answer the following fundamental question: is the event caused by the solar farm, i.e., is it locally-induced? or is it initiated in the grid, which consequently caused a response by the solar farm, i.e., is it grid-induced? Our answer to this question is based on an impedance-based method that is applied to the differential phasor representation of each event, coupled with a signature inspection method.
    
    \vspace{0.05cm}
    
    \item Regarding the locally-induced events, we seek to understand their engineering implications.
We observe that the majority of the locally-induced events happen during the low production periods of the solar farm. Furthermore, the events during the low production periods demonstrate more significant change in power factor. 

    \vspace{0.05cm}

\item Regarding the grid-induced events,
we characterize the response of the solar distribution feeder to such events. We also make comparisons with the response of an \emph{auxiliary neighboring feeder} to the same events. 

    \vspace{0.05cm}

\item We scrutinize multiple specific events that are particularly informative; such as by revealing the control system dynamics of the solar distribution feeder. The behavior of the solar farm is explained by the smart inverter control levels via dissecting two use cases.

\end{enumerate}

    \vspace{0.05cm}

The results in this study are insightful to utilities and solar power industry. They also provide new insight on the application of micro-PMU measurements in the study of behind-the-meter solar distribution feeders.

\vspace{0.1cm}

\subsection{Literature Review}

Power quality events that are associated with PV inverters in power distribution systems have been previously studied, such as in \cite{ seme2017power ,tonkoski2012impact, mahela2015detection, shaik2018power}. %
Some studies, such as in \cite{seme2017power}, use real-world measurements, while some others, such as in \cite{shaik2018power}, use computer simulations. More importantly, \emph{all} these prior studies have focused on typical \emph{load-serving feeders}, with varying PV penetration levels. In fact, to the best of our knowledge, this paper is the first data-driven event-based study of solar distribution feeders by using micro-PMU measurements. 

In terms of the relevant data-driven methodologies, machine learning techniques have been already used in \cite{wang2019detection, 6469207, 6553247, aligholian2019event} in order to detect power quality events in power distribution feeders. However, the previous studies have been concerned with load-serving distribution feeders. Therefore, while we used the same deep learning architecture as in \cite{aligholian2019event} for event detection in this paper; we had to \emph{train} the model with \emph{different data} from a real-world solar distribution feeder. 

Several studies have discussed identifying the source location of events in power distribution systems, e.g., in \cite{ biswal2016supervisory, sanitha2020micro, farajollahi2018locating }. Here, our concern is only on whether the source of the event is the solar farm itself; or the source of the event is the grid. In case of the latter, the solar farm still responds to the event.

\section{Background and Methodology}

The test site in this study is a solar distribution feeder that is dedicated to integrate about 200 PV inverters in a 4 MW solar farm into a distribution substation. The solar farm is behind-the-meter; however, a micro-PMU is available at the feeder-head at the distribution substation that provides us the voltage and current phasor measurements of this solar distribution feeder. This feeder does not have any load and all its solar power production is injected into the distribution substation.  

There is another micro-PMU that monitors a nearby feeder that contains a mix of major PV generation and major load. This \emph{auxiliary neighboring feeder} is sometimes a \emph{net load} and sometimes a \emph{net generator} during the period of our analysis. Our focus in this study is \emph{not} on this auxiliary neighboring feeder. However, in one part of our study, we use the synchronized micro-PMU measurements at this auxiliary feeder to better understand the behavior of the solar distribution feeder.

Next, we briefly overview the three key methodologies that we plan to use for our various analysis in this paper.

\subsection{Event Detection}\label{Detection}

The first step in our analysis is to 
detect the events from the micro-PMU measurements. This is a challenging task because events are inherently \textit{infrequent}, \textit{unscheduled}, and \textit{unknown}. Hence, there is no prior knowledge about their types and time of occurrence. Accordingly, in this paper, we used an unsupervised deep learning model from our previous work in \cite{aligholian2019event}, which implements Generative Adversarial Network (GAN) models. 
\color{black}
There are two main components in the event detection model; namely the \emph{generator} and and the \emph{discriminator}; which play a min-max game over the following function:

\begin{equation}
\begin{aligned}
V\left ( G,D \right ) = &\mathbb{E}_x\sim p_{data}(x)[log(D(x))] + \\
                        &\mathbb{E}_x \sim p_{z}(z)[log(1-D(G(z)))],
\end{aligned}
\label{eq:x1}
\end{equation}
\noindent where $V$ is the objective function, $G$ is the generator, $D$ is the discriminator, $p_{data}(x)$ is teh distribution of the real samples, and  $p_{z}(z)$ is the noise probability distribution function. 

For the GAN model, the optimal value of the min-max game over $V(G,D)$ in (\ref{eq:x1}) must satisfy the following two conditions:
\begin{itemize}
    \item \textbf{C1:} For any fixed $G$, the optimal discriminator {$D^*$} is:
    \begin{equation}
        D_G^*(x)=\frac{p_{data}(x)}{p_{data}(x)+p_{g}(x)}.
        \label{global1}
    \end{equation}
    \item \textbf{C2:} There exists a global solution such that: 
    \begin{equation}
        \begin{aligned}
             \min(\mathrel{\mathop{\max_{D}(V(G,D))}}) \Longleftrightarrow p_{g}(x)=p_{data}(x).
        \end{aligned}
        \label{global2}
    \end{equation}
\end{itemize}

\noindent where $p_{g}(x)$ is the distribution of the generated sample by the generator and $D_G^*$ is the optimal value for the output of the discriminator.

\color{black}

The model in \cite{aligholian2019event} was trained for load monitoring; therefore, we used the micro-PMU measurements from the solar distribution feeder to train new GAN models for the purpose of event detection at the solar distribution farm. See \cite{aligholian2019event} for more detail. 

The dataset under study consists of measurements from both PMUs for a period of ten days. A total of 229 events are detected at the solar distribution feeder; out of which 88 events are of interest because they happened during solar production. A total of 215 events are detected at the auxiliary neighboring feeder, which 87 of them happened during solar production of the solar distribution feeder.

\subsection{Event Region Identification}\label{region}
In reference to the location of the micro-PMU, an event occurs either in the \emph{upstream} or in the \emph{downstream} of the micro-PMU. The former is a \emph{locally-induced} event; while the latter is a \emph{grid-induced} event. To determine the source region of the event, we apply the following two different methods: 

\vspace{0.05cm}

\subsubsection{Impedance-based Method}\label{Zreal}
For each event, we can calculate the \emph{equivalent impedance}, denoted by $Z$, that is seen in the \emph{differential mode} in the upstream of the micro-PMU: 
\begin{equation} 
Z = \frac{\Delta V}{\Delta I} = \frac{V^\text{post} - V^\text{pre}}{I^\text{post} - I^\text{pre}}
\label{eq:impedance}
\end{equation}

\noindent where $I^\text{pre}$ and $V^\text{pre}$ are the current and voltage phasors that are seen in the steady-state condition right \emph{before} the event starts; and $I^\text{post}$ and $V^\text{post}$ are the current and voltage phasors that are seen during the steady state condition right \emph{after} the event settles down.
Both $\Delta V$ and $\Delta I$ are \emph{differential phasors} \cite{farajollahi2018locating, akrami2020sparse}. Our focus is on the resistive component of the $Z$. 
In particular, the event is considered to be \emph{locally-induced} if 
\begin{equation}
\operatorname{Real}\{Z\} > 0;
\label{eq:realz}
\end{equation}
otherwise, the event is considered to be \emph{grid-induced} \cite{ farajollahi2018locating}. We will use the impedance-based method in Section \ref{s2}.

\vspace{0.05cm}

\subsubsection{Comparison with Auxiliary Measurements} \label{combined}
This method takes advantage of the synchronized micro-PMU  measurements from the auxiliary neighboring feeder.
By comparing the signatures of an event on both feeders, i.e., the solar distribution feeder and the auxiliary neighboring feeder, we can determine that the event is grid-induced if it creates similar signatures on the voltage measurements on both feeders.
If the event is seen only on one feeder and there is no major signature on the other feeder, then it is a locally-induced event \cite{mohsenian2018distribution}. We will use the combination of the two methods in Section \ref{s3}.

\vspace{0.05cm}

\subsection{Event Dynamic}\label{control}

Some of the events that are captured in this study can reveal the dynamic behavior of the solar farm's control system.
\color{black}
The control systems of the inverters on a solar distribution feeder are highly convoluted. 
\color{black}
To dissect the event dynamics, we use the following four general control components \cite{guideline2018bps}:
\begin{itemize}
    \item Current regulation loop is the fastest loop that controls the injected currant by each inverter to the grid.  
    \item Voltage regulation loop is slower than the current regulation loop. It provides the setpoint for the current regulation loop; 
    upon changes in inverter terminal voltage.
    \item Maximum Power Point Tracking (MPPT) optimizes the utilization of input power for maximum power output.  
    For an individual inverter, this is the slowest controller. 
    \item Plant-level controller maintains the scheduled voltage and power factor (PF) of the system by coordinating the set points of individual inverter voltage or reactive power. Plant-level control speed is coordinated with the controls of individual inverters and is normally slower.
\end{itemize}

During an event, the \emph{major disturbances} are mostly controlled by faster controls; while \emph{minor disturbances} are mainly controlled by the plant-level controller response. Based on the above control components, two use cases are scrutinized in Section \ref{s5}.

\section{Analysis of Locally-Induced Events}\label{s2}
In this section, we study locally-induced events, make statistical conclusions, and discuss representative example events. 

\subsection{Event Correlation with PV Production Level}\label{II1}
Our analysis of the captured locally-induced events reveals a relationship between event occurrence and the solar production level. In particular, we have observed that the majority of the locally-induced events occur during low production period. 

It is worth clarifying that what we refer to as event in the micro-PMU data is very different from the regular fluctuations in the solar production level that are due to the changes in solar irradiance. This point is explained in Fig. 1. 
Here, we show two example events, denoted by Event 1 in Fig. 1(a) and Event 2 in Fig. 1(b). These two events are much \emph{smaller in magnitude} and much \emph{shorter in duration} compared to the typical fluctuation in solar production level. 
\color{black}
These kinds of events are captured only by the installed micro-PMU. They cannot be captured by regular meters. The solar production level is 9.2\% during Event 1 and 14.8\% during Event 2.
\color{black}

Fig.~\ref{fig1} shows the scatter plot for all the captured events that are identified as locally-induced. Here the focus is on the events with $\text{Real}\{Z\} > 0$, see the methodology in Section \ref{Zreal}. There is an inverse correlation between the production level and the number of events, which can be expressed as an exponential decay function, as presented in  \eqref{eq:curve1} and Fig.~\ref{fig1}(a). 
\color{black}
\begin{equation}
y = a.x^{b}+c,
\label{eq:curve1}
\end{equation}

\noindent where $x$ and $y$ are the production level and the real part of the impedance $Z$ that we defined in (\ref{eq:impedance}). Parameters $a$, $b$, and $c$ are obtained through curve fitting as $850$, $-1$, and $50$, respectively.  

\color{black}

\begin{figure}[tbp]  
\centerline{\includegraphics[width=90mm ,scale=1]{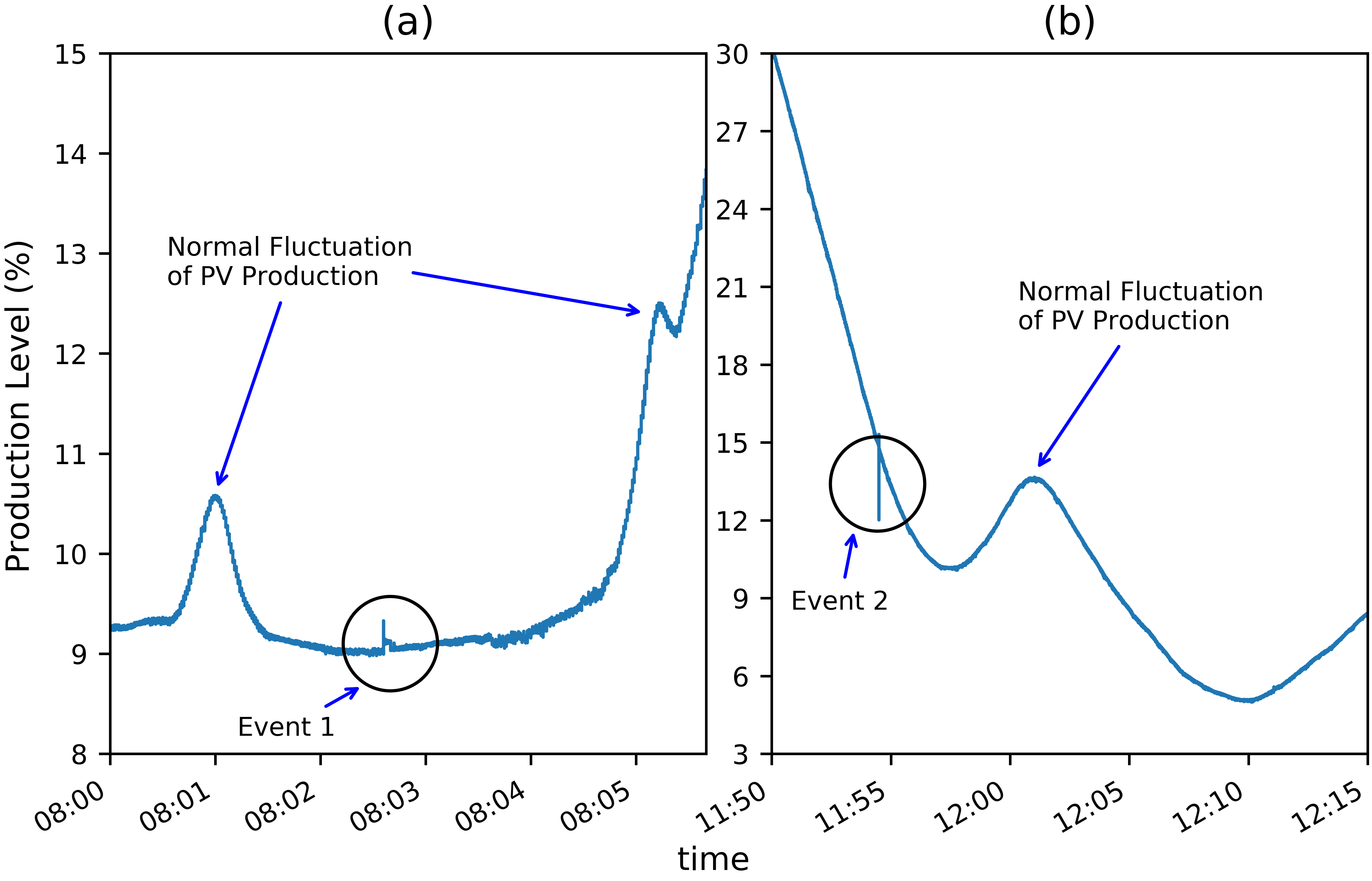}}
\vspace*{-2.5mm}
\caption{Two sample locally-induced events: (a) Event 1 shortly after production starts; (b) Event 2 occurs around noon on a cloudy day.}
\label{fig:lowprodevents}
\vspace*{-1mm}
\end{figure}

As shown in Fig.~\ref{fig1}(b), about 70\% of the locally-induced events happened when the solar production was at \%30 or less. That means, either more control actions took place at the solar farm during low production periods; or the control actions are more impactful during such periods and therefore their signatures are more visible. In either case, these results highlight the importance of monitoring the operation of the solar distribution feeder during low production periods.  

\begin{figure}[t]
\centerline{\includegraphics[width=90mm ,scale=1]{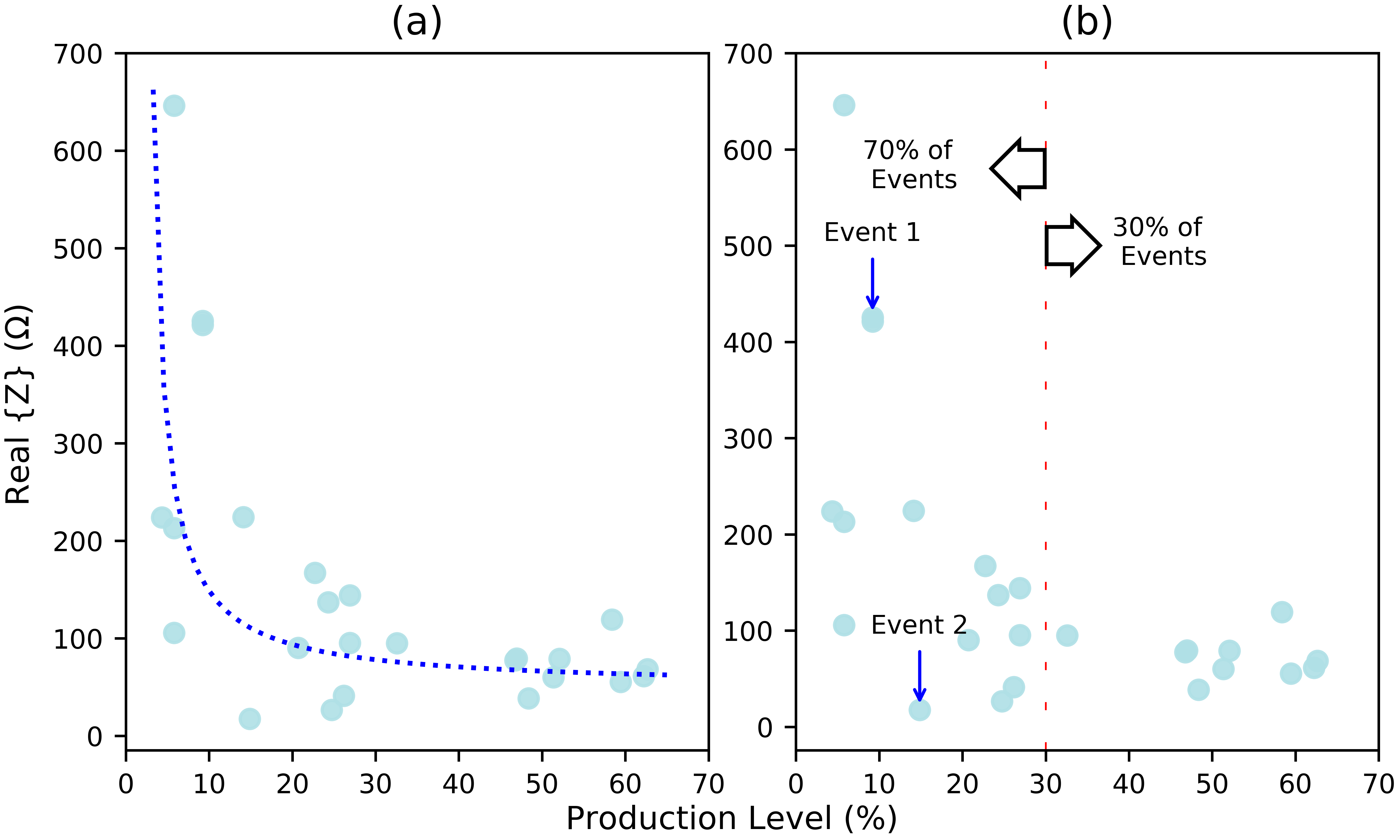}}
\vspace*{-2.5mm}
\caption{Locally-induced events on the solar distribution feeder: (a)  exponential decay relation between production level and number of events; (b) inverse correlation between the PV production level and the number of events.}
\label{fig1}
\end{figure}

\begin{figure}[t]  
\centerline{\includegraphics[width=90mm ,height=55.1mm]{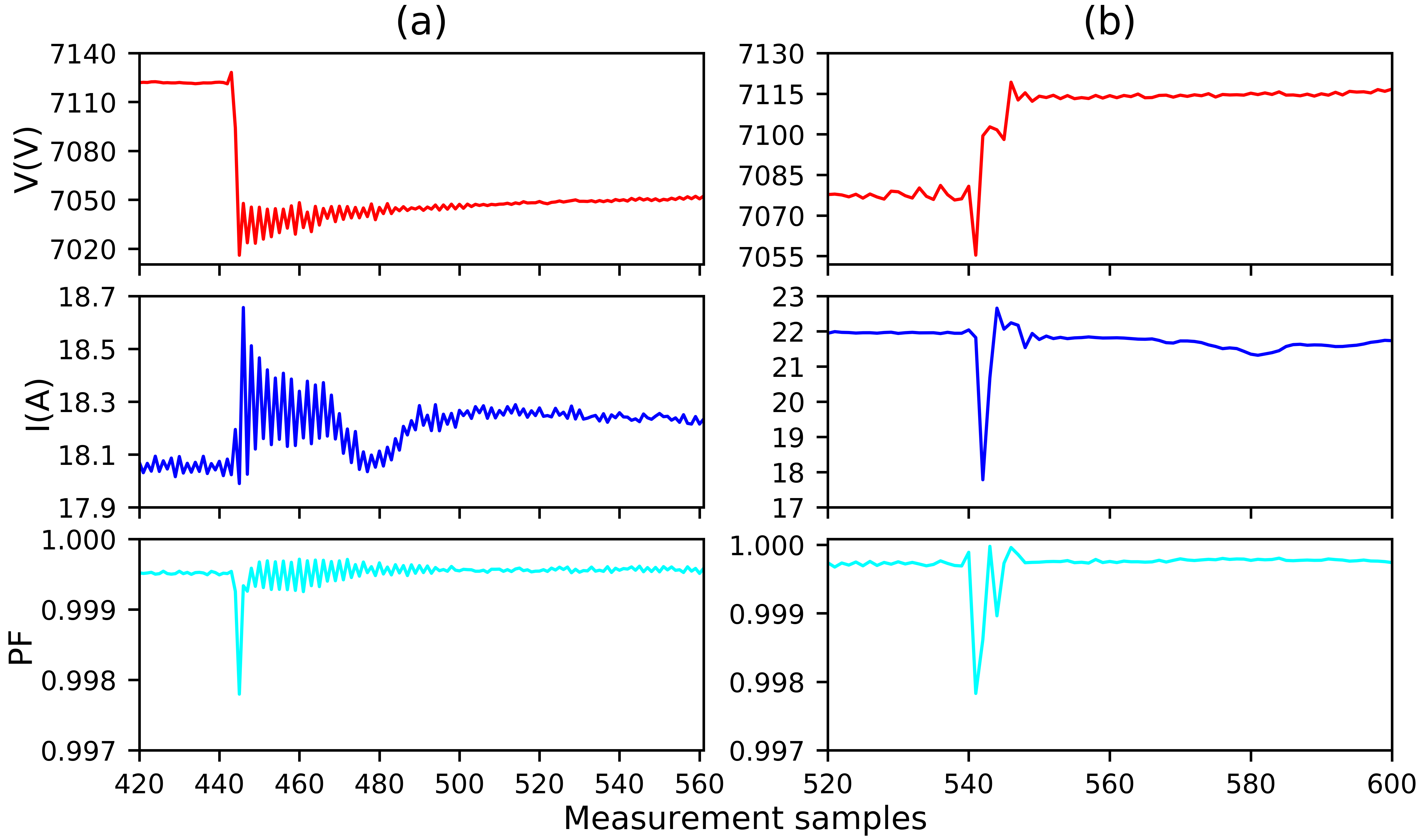}}
\vspace*{-2.5mm}
\caption{High resolution event signatures: (a) Event 1; and (b) Event 2.  }
\vspace*{-5mm}
\label{fig:twoevents}
\end{figure}

A closer look of Events 1 and 2 is provided in Fig.~\ref{fig:twoevents}. For Event 1 in Fig.~\ref{fig:twoevents}(a), we can obtain $\Delta{V}$ $= 238.6+ 1789.5 j$ and $\Delta{I}$ $= -1.835 - 1.934 j$. From \eqref{eq:impedance}, we have:
$\operatorname{Real}\{Z\} = 425.37$. Therefore, from \eqref{eq:realz}, we can conclude that Event 1 is a locally-induced event. 
For Event 2 in Fig.~\ref{fig:twoevents}(b), we can obtain $\Delta{V} = 2.8 -29.5 j$ and $\Delta{I} = -0.363 - 1.561 j$. From \eqref{eq:impedance}, we have:  
$\operatorname{Real}\{Z\} =  17.57$. Thus, from \eqref{eq:realz}, we can conclude that Event 2 is also a locally-induced event.

\subsection{Changes in Power Factor}\label{II2}

\begin{figure}[t]  
\centerline{\includegraphics[width=90mm ,height=55.1mm]{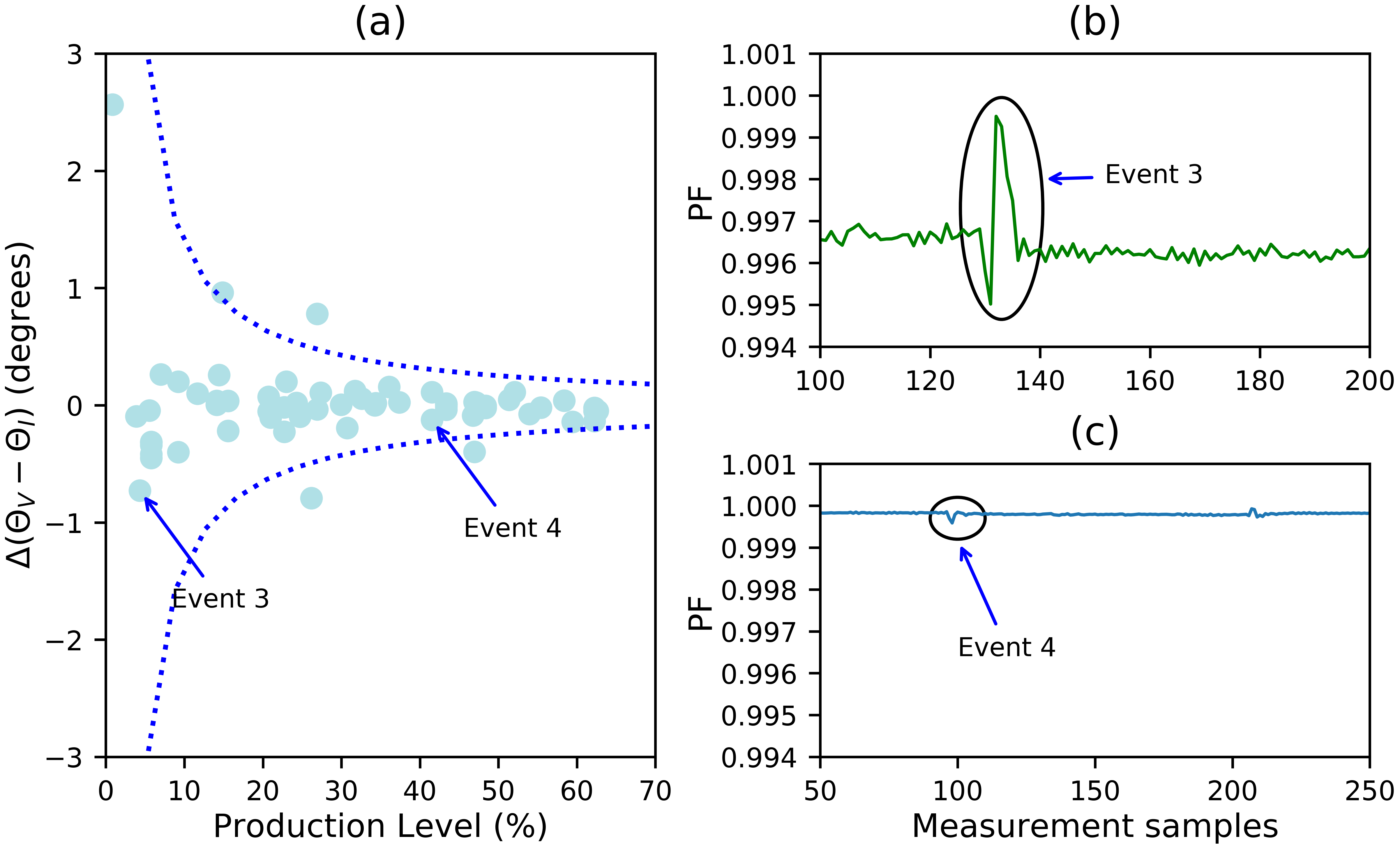}}
\vspace*{-2.0mm}
\caption{(a) Relationship between disturbance in phase-angle differences and production level;
(b) small change in PF in Event 3 during low  production; and (c) almost no change in PF in Event 4 during high production.}
\vspace{-0.2cm}
\label{fig_angle_PF}
\vspace*{-1.3mm}
\end{figure}

We further observed that, not only the majority of the locally-induced events occurred during low production periods, but also the events that occurred during low production periods demonstrated more significant changes in power factor (PF). In Fig.~\ref{fig_angle_PF}(a), we can see the change in the \emph{phase angle difference} between voltage and current phasors that are caused by the locally-induced events versus the production level. 

Note that, the cosine of the quantity in the y-axis provides the \emph{change} in power factor. Here, $\theta_V$ and $\theta_I$ denote the phase angle measurements in voltage and in current, respectively. The changes that caused by the events in the phase angle differences \emph{declines exponentially} with the increase in production level. See the formulation in \eqref{eq:curve2} and Fig.~\ref{fig_angle_PF}(a).
\color{black}
\begin{equation}
y = d.x^{e}
\label{eq:curve2}
\end{equation}
where parameters $d$ and $e$ are obtained through curve fitting. Parameter $d$ is $4.5$ and $-4.5$ for the exponential analysis and its inverse, respectively. Parameter $e$ is obtained as $-1$.

\color{black}

Next, we look at two events which took place at different production levels and led to different changes in phase angle differences, and accordingly in PF. In Fig.~\ref{fig_angle_PF}(b), Event 3 happened at a low production level and caused a \emph{major} change in PF, however in Fig.~\ref{fig_angle_PF}(c), Event 4 happened at a higher production level and resulted in an insignificant PF agitation.

 The critical low production periods are the loss of voltage support at sunset, particularly at large PV penetration, during \emph{startup} and \emph{shutdown} of the inverters. Events during these periods are impacting the power system more, by affecting the phase angle and consequently the power factor of the system.

\section{Analysis of the Response \\ to Grid-Induced Events}\label{s3}

Since we have access to the \emph{synchronized} micro-PMU measurements at the auxiliary neighboring feeder, we can also apply the \emph{signature inspection} method, described in Section \ref{region}, to see if a grid-induced event also creates a signature on the voltage phasor measurements at the auxiliary neighboring feeder. 

Among all 88 events associated with the solar distribution feeder, only 8 events are recognized as grid induced events based on the $\text{Real}\{Z\} < 0$ criteria. Interestingly, \emph{all} these eight events create signature also on the auxiliary neighboring feeder. Therefore, they are confirmed with \emph{both} methods to be grid-induced events. Moreover, these observations verify the reliability of the impedance-based method. 

\begin{figure}[t]  
\centerline{\includegraphics[trim=0 0 0 0,clip,width=90mm ,height=55.1mm]{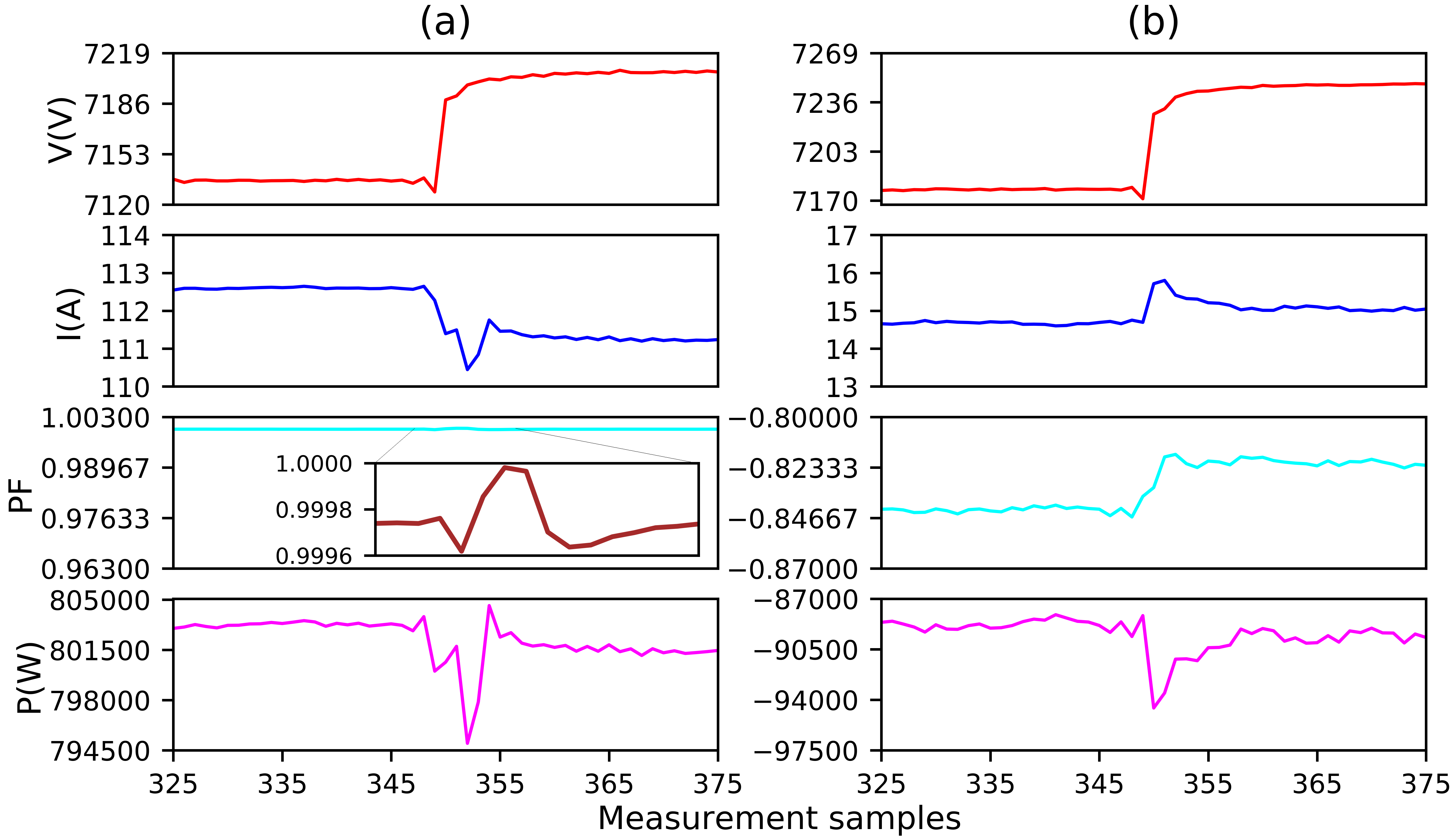}}
\vspace*{-2.5mm}
\caption{Response of the two feeders to grid-induced event: (a) response of the solar distribution feeder; (b) response of the auxiliary neighboring feeder.}
\vspace*{-1.2mm}
\label{fig43}
\end{figure}

The response of the two feeders to an example grid-induced event is shown in Fig.~\ref{fig43}. Let us refer to this event as Event 5. Notice how the voltage suddenly increases on both set of  measurements. The two responses have major differences. In particular, the event causes only a \emph{very minor transient} change in the power factor in Fig.~\ref{fig43}(a). However, the same event causes a \emph{major steady} change in the power factor in Fig.~\ref{fig43}(b). The change in the magnitude of current are also in the opposite directions in Figs.~\ref{fig43}(a) and (b). It should be noted that the auxiliary neighboring feeder is a net load during this event. The behavior of the solar distribution feeder can be potentially attributed to the MPPT behavior of the PV inverters. As for the auxiliary neighboring feeder, it appears to act as a constant-impedance load; while its power factor is also affected.

\section{Analysis of Event Dynamics \\ with PV Inverters Operation}\label{s5}

As discussed in Section \ref{control}, the control system in a solar distribution feeder has multiple control loops. Therefore, the dynamic response of the solar distribution feeder to a grid-induced event can constitute various stages. 

\begin{figure}[t]  
\centerline{\includegraphics[width=90mm ,height=55.1mm]{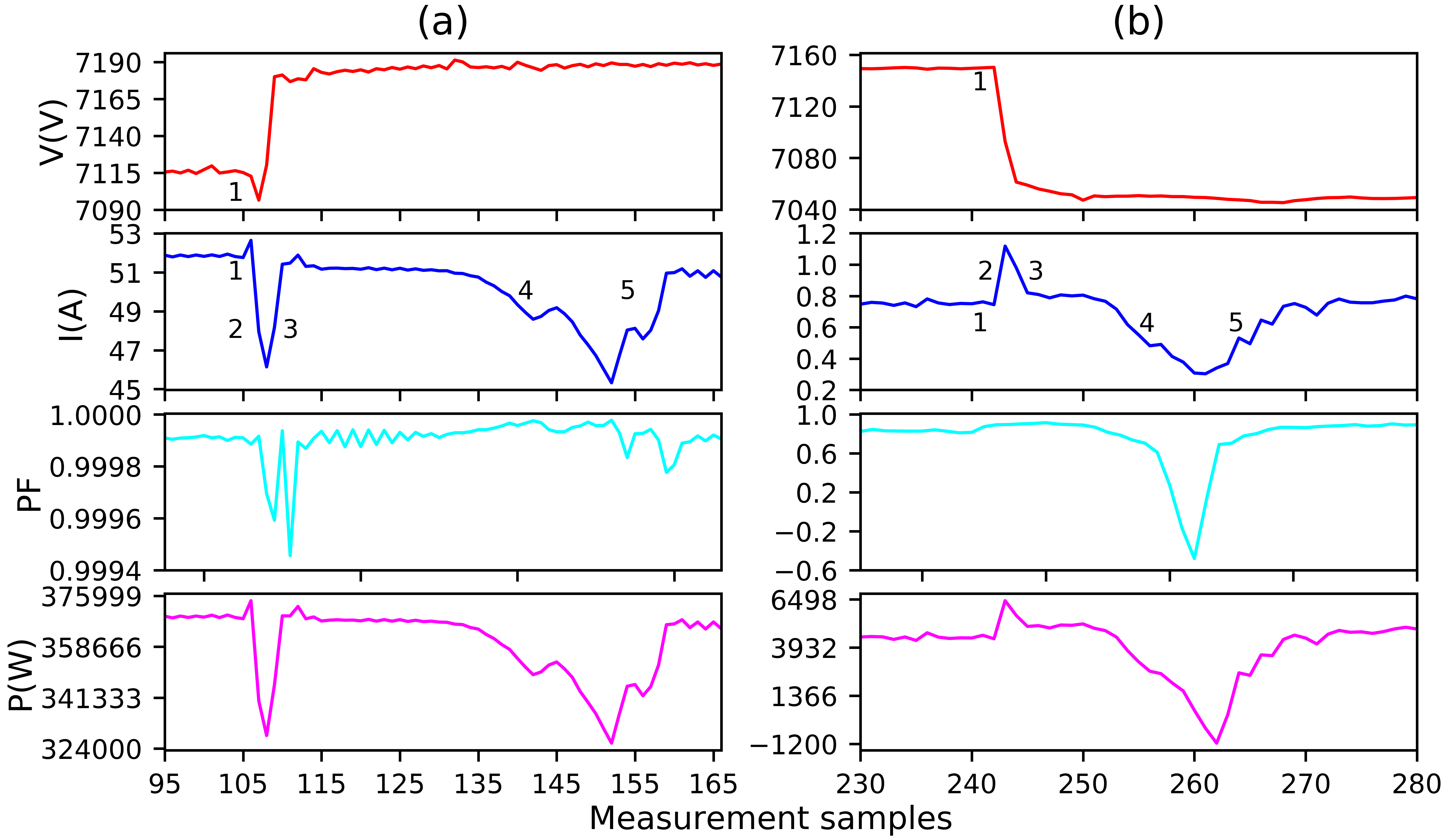}}
\vspace*{-2.5mm}
\caption{Comparing and characterizing the dynamic response of the solar distribution feeder to two different grid-induced events:  (a) Event 6, which is a step-up change in voltage; and (b) Event 7; which is a step-down in voltage.}
\label{dynaimc}
\vspace*{-1.3mm}
\end{figure}

In this section, we examine the dynamic response of the solar distribution feeder to two distinct grid-induced events;  as shown in Fig.~\ref{dynaimc}(a) and Fig.~\ref{dynaimc}(b). 
We refer to these events as Event 6 and Event 7, respectively.
The response of the solar distribution feeder to each event is broken down into \emph{five stages} that are marked from 1 through 5 on each graph. 

  In Stage 1, the event occurs. Event 6 is a  sudden rise voltage; see Fig.~\ref{dynaimc}(a).  is a sudden drop in voltage; see Fig.~\ref{dynaimc}(b); thus, triggering a response by the PV control system due to detecting the changes at the inverter voltage terminals.
  In Stage 2, the immediate reaction of the system is to keep the output power stable; this causes a prompt current drop in Fig.~\ref{dynaimc}(a) to decrease the dc-bus voltage; and a prompt current rise in Fig.~\ref{dynaimc}(b) to increase the dc-bus voltage.
  In Stage 3, the new dc-bus voltage level modifies the MPPT output; which is the input to the voltage regulation loop. Subsequently, the reference is fine-tuned for the current regulation loop. As a result, current increases in Event 6 and decreases in Event 7. Thus, the dc-bus voltage goes back to its pre-disturbance value.
  In Stage 4, after passing the initial transient conditions, the plant level control regulates the process and applies the ramp rate limitation. This results in a momentary decrease of the current on both events.
  Finally, in Stage 5, by a moderate ramp rate, the plant level controller brings back the current to the regulated set point.

\section{Conclusions and Future Work}

An event-based analysis is conducted based on real-world micro-PMU measurements at a solar distribution feeder. After detecting all the events by training an unsupervised deep learning model, the events are classified based on their origin; either locally-induced by the solar farm itself; or they are grid-induced. 
\color{black}
It was observed that 70\% of the events happen when the solar production is at \%30 or less.
\color{black}
Furthermore, the  events  during  the  low solar production periods demonstrate more significant changes in power factor.
\color{black}
In low production, the range of the phase angle change is $\pm3^{\circ}$, while in high production, it is mostly around $0^{\circ}$.  

Among the 88 events that were detected at the solar distribution feeder,  only  8  events were  recognized  as  grid-induced.\color{black} 
 We also examined the response of the solar distribution feeder to all these grid-induced events; and compared it to the response of a neighboring feeder to the exact same events. Finally, we characterized the event dynamics \color{black}into 5 steps \color{black} based on the control system mechanisms of the solar distribution feeder.

The analysis in this paper provides awareness about the operation of solar distribution feeders. In the future, the results and findings in this paper can be used in signature mapping for the purpose of diagnostics and prognostics application in higher-penetration solar distribution feeders. 

\color{black}
This study is also beneficial in monitoring the health of equipment in the solar farm. Therefore, it is of economic value to the utility due to providing with insight on the state of the health of the inaccessible behind-the-meter solar farm equipment. Importantly, this type of study is also essential to address the cascading effect of ongoing solar energy increment on the stability and operation of the distribution systems. 
\color{black}

\bibliography{main} 
\bibliographystyle{ieeetr}

\vspace{12pt}

\end{document}